\documentclass[aps,prd,superscriptaddress,amsfonts,
showpacs,preprintnumbers,nofootinbib,onecolumn,a4]{revtex4}
\usepackage{graphicx,color,here}
\newcommand{\bfk}{\mbox{\boldmath $k$}}
\newcommand{\bfp}{\mbox{\boldmath $p$}}
\newcommand{\bfP}{\mbox{\boldmath $P$}}
\begin{document}

\title{Transverse momentum dependence of the quark helicity distributions and
the Cahn effect in double-spin asymmetry $A_{LL}$ in Semi Inclusive DIS}
\author {{\bf M.~Anselmino$^{1}$, A.~Efremov$^{2}$,
A.~Kotzinian$^{2,3,4}$ and B.~Parsamyan$^{3}$}\\
\vskip 0.2cm
{\it $^1$ Dipartimento di Fisica Teorica, Universit\`a di Torino, and \\
INFN, Sezione di Torino, Via P. Giuria 1, I-10125 Torino, Italy }\\
\vskip 4pt
{\it $^2$\it JINR, 141980 Dubna, Russia}\\
\vskip 4pt
{\it $^3$ Dipartimento di Fisica Generale, Universit\`a di Torino, and \\
INFN, Sezione di Torino, Via P. Giuria 1, I-10125 Torino, Italy}\\
\vskip 4pt
{\it $^4$ Yerevan Physics Institute, 375036 Yerevan, Armenia} \\
}

\begin{abstract}
Within the LO QCD parton model of Semi Inclusive Deep Inelastic Scattering,
$\ell \, N \to \ell \, h \, X$, with unintegrated quark distribution and
fragmentation functions, we study the $\bfP_{hT}$ dependence of the double
longitudinal-spin asymmetry $A_{LL}$. We include $1/Q$ kinematic corrections,
which induce an azimuthal modulation of the asymmetry, analogous to the Cahn
effect in unpolarized SIDIS. We show that a study of $A_{LL}$ and of the
weighted DSA $A_{LL}^{\cos\phi_h}$ allows to extract the transverse momentum
dependence of the unintegrated helicity distribution function
$g_{1L}^q(x,k_\perp)$ [or $\Delta q(x, k_\perp)]$. Predictions, based on some
models for the unknown functions, are given for ongoing COMPASS, HERMES and
JLab experiments.
\end{abstract}
\maketitle
%\newpage
\section{\label{sec:intro}Introduction}

After many years of intensive experimental and theoretical study the partonic
origin of the nucleon spin still remains mysterious. In particular, while a
lot of understanding has been achieved concerning the longitudinal structure
of a fast moving proton -- the $x$-dependence of the parton distribution
functions and of the helicity distributions -- very little is known about the
transverse structure. Transverse refers to the direction of motion and
concerns both the transverse spin distributions and the parton intrinsic
motion, $\bfk_\perp$. These unexplored degrees of freedom cannot be considered
as minor details in our modeling of the nucleon. Without a good knowledge of
the total intrinsic momentum carried by the partons, and its connection with
the spin, we could never understand the parton orbital motion and progress
towards a more structured picture which goes beyond the simple collinear
partonic representation.

Recent data on single spin azimuthal asymmetries in semi-inclusive deep
inelastic scattering (SIDIS, $\ell \, N \to \ell \, h \, X$) obtained by
HERMES~\cite{herm}, COMPASS~\cite{comp} and CLAS--JLab~\cite{jlab}
collaborations triggered a lot of interest towards the transverse momentum
dependent (TMD) and spin dependent distribution and fragmentation functions
(PDFs and FFs). In fact spin-$\bfk_\perp$ correlations induce new spin
effects, which would be zero in the absence of intrinsic motion and are
related to transverse polarizations. On the other hand, it has been known
for a long time that the dependence on the transverse final hadron momentum,
$\bfP_{hT}$, observed in unpolarized SIDIS processes, can be related, for
$P_{hT} = |\bfP_{hT}|$ values up to about 1 GeV/$c$, to quark intrinsic
motion \cite{cahn,anskpr,noi2}.

We consider here polarized SIDIS processes, at twist-two in the parton model,
with transverse momentum dependent distribution and fragmentation functions.
Such processes can be described in terms of six time reversal
even~\cite{ko, mt} and two (na\"{\i}vely) time reversal odd PDFs.
The dependence on partonic intrinsic motion induces a dependence on $P_{hT}$.
In addition, at $\mathcal{O}(k_\perp/Q)$, kinematic corrections induce a
dependence of the unpolarized cross section on the azimuthal angle $\phi_h$
between the leptonic and the hadron production planes -- the so called Cahn
effect~\cite{cahn}. It was shown in Ref.~\cite{anskpr} that a careful study of
the dependence of the cross section on the final hadron momentum allows to
extract the average values of intrinsic momenta in unpolarized PDFs and FFs.

We expand on the work of Ref.~\cite{anskpr} and evaluate the role of partonic
intrinsic motion in polarized SIDIS; in particular, on the double spin
asymmetry (DSA) for the scattering of longitudinally polarized leptons off a
longitudinally polarized target, $A_{LL}$, where longitudinal refers to the
incoming lepton direction, in the laboratory frame. We show that a study of
$A_{LL}$ and of the weighted asymmetry $A_{LL}^{\cos\phi_h}$ allows to
learn about the $k_\perp$ dependence of the quark helicity distribution
$g_{1L}^q(x, k_\perp)$ [or $\Delta q(x, k_\perp)]$. Similar results for the
TMD DF $g_{1T}^q$ have been obtained in a recent paper~\cite{kpp}, by
considering the longitudinal--transverse DSA $A_{LT}^{\cos(\phi_h-\phi_S)}$.

The article is organized as follows. In Sec~\ref{sec:xsec} we shortly recall
the relevant formalism for polarized SIDIS. In Sec~\ref{sec:asym} some
predictions for the double longitudinal spin asymmetries are presented.
The results are given for different sets of kinematical cuts, according to
the experimental setups of HERMES, COMPASS and JLab experiments; they indicate
the best kinematical regions for the asymmetry to be sizeable.
Finally, in Sec.~\ref{sec:concl} we shortly discuss our results and draw some
conclusions.

\section{Polarized cross section \label{sec:xsec}}

Following Ref.~\cite{ko}, we consider the polarized SIDIS in the
simple quark-parton model, with unintegrated parton distributions.
We use the standard notations for DIS variables: $\ell$ and $\ell'$
are, respectively, the four-momenta of the initial and the final
state leptons; $q = \ell - \ell'$ is the exchanged virtual photon
momentum; $P$ ($M$) is the target nucleon momentum (mass), $S$ its
polarization; $P_h$ is the final hadron momentum; $Q^2=-q^2$;
$x=Q^2/2P\cdot q$; $y=P\cdot q/P\cdot \ell$; $z=P\cdot P_h/P\cdot
q$, $Q^2 = xy(s - M^2)$, $s = (\ell + P)^2$. We work in a frame with
the $z$-axis along the virtual photon momentum direction and the
$x$-axis in the lepton scattering plane, with positive direction
chosen along the lepton transverse momentum. The produced hadron has
transverse momentum $\bfP_{hT}$; its azimuthal angle, $\phi_h$, and
the azimuthal angle of the transverse nucleon spin, $\phi_S$, are
measured around the $z$-axis (for further details see Ref.
\cite{ko}).

We consider longitudinally polarized protons and leptons, where longitudinal
(according to the laboratory setup) refers to the initial lepton direction. It
then results that a proton with longitudinal spin $S$ along the incoming
lepton direction, has a transverse -- with respect to the $\gamma^*$
direction -- spin component:
\begin{equation}\label{sintg}
    S_T = S\sin\theta_\gamma,
\end{equation}
where
\begin{equation}\label{ttrpol}
\sin\theta_\gamma = \sqrt{\frac{4M^2x^2}{Q^2+4M^2x^2}
\left( 1-y - \frac{M^2x^2y^2}{Q^2} \right)}
\simeq \frac{2Mx\sqrt{1-y}}{Q}\;\cdot
\end{equation}
This component gives contributions of order $M/Q$.

Keeping only twist-two contributions and terms up to $\mathcal{O}(M/Q)$ the
cross section for SIDIS of longitudinally polarized leptons off a
longitudinally polarized target can be written as:
\begin{equation}\label{dsll}
\frac{d^5\sigma^{\begin{array}{c}\hspace*{-0.1cm}\to\vspace*{-0.25cm}\\
\hspace*{-0.1cm}\Leftarrow\end{array}}}{dx \, dy \, dz \, d^2P_{hT}} =
\frac{2 \alpha^2}{x y^2 s}\, \left\{ {\cal H}_{f_1} + \lambda \, (S_L
{\cal H}_{g_{1L}} + S_T{\cal H}_{g_{1T}}) \right\}, \label{sig}
\end{equation}
where the arrows indicate the direction of the lepton
($\rightarrow$) and target nucleon ($\Leftarrow$) polarizations,
with respect to the lepton momentum; $\lambda$, $S_L$ and $S_T$ are
the magnitudes of, respectively: the longitudinal beam polarization,
the longitudinal and the transverse target polarization. Notice that
$\Leftarrow$ stands for a nucleon with a polarization vector, in the
laboratory frame where the nucleon is at rest, \textit{opposite} to
the initial lepton momentum. For a $\Rightarrow$ polarization one
reverses the signs of the $S_L$ and $S_T$ terms.

The three terms have a simple partonic interpretation:
\begin{equation}\label{hf1p}
{\cal H}_{f_1} = \sum_q e_q^2 \int d^2\bfk_\perp \, f^q_1(x,k_\perp) \, \pi
y^2 \, \frac{\hat s^2 + \hat u^2} {Q^4} \, D_q^h(z, p_\perp),
\end{equation}
\begin{equation}\label{hg1lp}
{\cal H}_{g_{1L}} = \sum_q e_q^2 \int d^2\bfk_\perp \, g^q_{1L}(x, k_\perp) \,
\pi y^2 \, \frac{\hat s^2 - \hat u^2} {Q^4} \, D_q^h(z, p_\perp),
\end{equation}
\begin{equation}\label{hg1tp}
{\cal H}_{g_{1T}} = - \sum_q e_q^2 \int d^2\bfk_\perp \, \frac{k_\perp}{M} \,
\cos\varphi \> g^{q\perp}_{1T}(x, k_\perp) \, \pi y^2 \, \frac{\hat s^2 - \hat
u^2} {Q^4} \, D_q^h(z, p_\perp) \>,
\end{equation}
and deserve some comments.
\begin{itemize}
\item
The partonic factorized structure of the above equations is supposed to hold
in the large $Q^2$ kinematic region where
$P_{hT} \simeq \Lambda_{\rm{QCD}}\simeq k_\perp \ll Q$ \cite{jmy}.
It neglects terms of $\mathcal{O}(k_\perp/Q)^2$, in which case
$$\bfp_{\perp} = \bfP_{hT} - z \bfk_\perp \>,$$
where $\bfp_\perp$ is the intrinsic transverse momentum of the hadron $h$ with
respect to the fragmenting quark direction.
\item
The first two contributions, Eqs. (\ref{hf1p}) and (\ref{hg1lp}), give,
respectively, the unpolarized cross section and the helicity asymmetry
\begin{equation}
\label{2terms}
\frac{d^5\sigma}{dx \, dy \, dz \, d^2 P_{hT}} = \frac{2 \alpha^2}
{x \, y^2s} \> {\cal H}_{f_1} \quad\quad\quad\quad
\frac{d^5\sigma^{++}} {dx \, dy \, dz \, d^2 P_{hT}} -
\frac{d^5\sigma^{+-}} {dx \, dy \, dz \, d^2 P_{hT}} =
\frac{4 \alpha^2} {x \, y^2s} \> {\cal H}_{g_{1L}} \>,
\end{equation}
where $+,-$ stand for helicity states. The quark intrinsic motion
induces a \textit{kinematical azimuthal dependence}, via the
elementary polarized cross section~\cite{ko}
\begin{equation}\label{lq}
\frac{d\sigma^{lq \rightarrow lq}}{dQ^2 \, d\varphi} \propto
\frac{{\hat s}^2 + {\hat u}^2 + \lambda \, \lambda_q \,
({\hat s}^2 - {\hat u}^2)}{{\hat t}^2},
\end{equation}
where $\lambda_q$ denotes the quark helicity. Keeping the terms up
to order of $k_\perp/Q$ the Mandelstam variables for the
non-coplanar $\ell q \rightarrow \ell q$ scattering are expressed as
\begin{eqnarray}\label{mand}
  \nonumber {\hat s}&\simeq& xs \left[1-2\sqrt{1-y} \>\, \frac{k_\perp}{Q} \>
  \cos\varphi \right], \\
  {\hat t}&=& -Q^2 = -xys, \\
  \nonumber {\hat u}&\simeq& -xs \,(1-y) \left[1-\frac{2 \, k_\perp}
  {Q \, \sqrt{1-y}}\, \cos\varphi\right],
\end{eqnarray}
where $\varphi$ is the azimuthal angle of $\bfk_\perp$,
$d^2\bfk_\perp = k_\perp \, dk_\perp \, d\varphi$. Eq. (\ref{hf1p})
then gives the unpolarized Cahn effect \cite{cahn}, while Eq.
(\ref{hg1lp}) gives the corresponding effect for the polarized
(helicity) cross section, both at $\mathcal{O}(k_\perp/Q)$.
\item
Eq. (\ref{hg1tp}) contains another $\cos\varphi$ dependence, of different
origin. While the distribution functions $f^q_1(x,k_\perp)$ and
$g^{q}_{1L}(x,k_\perp)$ which appear in Eqs. (\ref{hf1p}) and (\ref{hg1lp}),
are just the $k_\perp$ dependent unpolarized and longitudinally polarized
(helicity) PDFs, which, upon integration over $d^2\bfk_\perp$, give the usual
$f_1^q(x)$ [or $q(x)$] and $g_1^q(x)$ [or $\Delta q(x)$] distributions, the
quantity
\begin{equation}\label{delta+ST}
- \frac{k_\perp}{M} \, \cos\varphi \> g^{q\perp}_{1T}(x,k_\perp) = \Delta\hat
f_{s_z/S_T}
\end{equation}
is related to the number of partons longitudinally polarized inside
a transversely polarized proton \cite{ko,mt,noi}: it can only depend
on the scalar product between the two corresponding polarization
vectors, which gives the $\cos(\phi_{S_T} - \varphi) = -
\cos\varphi$ factor explicitly shown (see, for example, Eq. (C19) of
Ref. \cite{noi}). This distribution is a leading-twist one, not
suppressed by $(k_\perp/Q)$ small factors. However, Eq.
(\ref{hg1tp}) will be multiplied by $S_T$, which is of
$\mathcal{O}(M/Q)$, Eqs. (\ref{sintg})--(\ref{dsll}); for this
reason, in Eq. (\ref{hg1tp}) we shall not take into account the
extra $(k_\perp/Q)$ kinematical terms contained in $(\hat s^2 -\hat
u^2)$ of Eq. (\ref{lq}).
\end{itemize}

The integrals in Eqs. (\ref{hf1p})--(\ref{hg1tp}) can be
analytically performed, if one assumes a simple factorized and
gaussian behaviour of the involved TMD PDFs and FFs
\begin{eqnarray}
f^q_1(x,k_\perp) &=& f^q_1(x) \, \frac{1}{\pi \mu_0^2}\,
\exp\left( -\frac{k_\perp^2}{\mu_0^2} \right)\label{dfffg1},\\
D_q^h(z, p_\perp)&=& D_q^h(z) \, \frac{1}{\pi \mu_D^2}\,
\exp\left( -\frac{p_\perp^2}{\mu_D^2} \right)\label{dfffg2},\\
g^{q\perp}_{1T}(x,k_\perp)&=& g^{q}_{1T}(x) \, \frac{1}{\pi \mu_1^2}\,
\exp\left( -\frac{k_\perp^2}{\mu_1^2} \right),\label{dfffg3}\\
g^q_{1L}(x,k_\perp)&=& g^q_{1}(x) \, \frac{1}{\pi \mu_2^2}\,
\exp\left( -\frac{k_\perp^2}{\mu_2^2} \right),\label{dfffg4}
\end{eqnarray}
yielding, at $\mathcal{O}(P_{hT}/Q)$:
\begin{equation}\label{hf1}
{\cal H}_{f_1} = \left[ 1 + (1 - y)^2  - 4(2 - y)\, \sqrt {1 - y} \>
\frac{z \, \mu_0^2 \, P_{hT}} {Q \, (\mu_D^2 + z^2 \mu_0^2)} \,
\cos \phi _h \right]
\frac{\exp\left(-\frac{P_{hT}^2}{\mu_D^2 + z^2 \mu_0^2} \right)}
{\mu_D^2 + z^2 \, \mu_0^2} \sum_q e_q^2\,f^q_1(x)\,D_q^h(z),
\end{equation}
\begin{equation}\label{hg1l}
{\cal H}_{g_{1L}} = y \left[ 2 - y - 4 \, \sqrt {1 - y} \,
\frac{z \, \mu _2^2 \, P_{hT}} {Q \, (\mu_D^2 + z^2 \mu_2^2)} \,
\cos \phi_h \right]
\frac{\exp\left(-\frac{P_{hT}^2}{\mu_D^2 + z^2 \mu_2^2}\right)}
{\mu_D^2 + z^2 \, \mu_2^2} \sum_q e_q^2\,g^q_1(x)\,D_q^h(z),
\end{equation}
\begin{equation}\label{hg1t}
{\cal H}_{g_{1T}} = -y(2-y) \, \frac{z \, \mu_1^2 \, P_{hT}}
{M (\mu_D^2 + z^2 \mu_1^2)} \cos \phi _h \>
\frac{\exp\left( -\frac{P_{hT}^2}
{\mu_D^2 + z^2 \mu_1^2}\right)} {\mu_D^2 + z^2 \mu_1^2}
\sum_q e_q^2\,g^{q}_{1T}(x)\,D_q^h(z) \>.
\end{equation}

\section{Predictions for $A_{LL}$ \label{sec:asym}}

We use Eqs. (\ref{sig}) and (\ref{hf1})--(\ref{hg1t}) to compute observables
which depend on partonic intrinsic motions. Notice that we have allowed
different average values of $\langle k_\perp^2\rangle$ for the different
distribution functions: $\langle k_\perp^2\rangle = \mu_0^2$ for the
unpolarized distributions, $\langle k_\perp^2\rangle = \mu_2^2$ for the
helicity distributions, and $\langle k_\perp^2\rangle = \mu_1^2$ for
$g^{q\perp}_{1T}(x,k_\perp)$; each of these value is taken to be constant
and flavour independent. For the fragmentation functions we have
$\langle p_\perp^2\rangle = \mu_D^2$. Following Ref. \cite{anskpr} we use
\begin{equation}\label{par}
\mu_0^2 = 0.25 \> ({\rm GeV}/c)^2 \quad\quad\quad\quad
\mu_D^2 = 0.20 \> ({\rm GeV}/c)^2 \>,
\end{equation}
while we consider $\mu_1^2$ and $\mu_2^2$ as free parameters, which
can give interesting information on the quark transverse motion in
polarized protons; the na\"{\i}ve positivity bounds imply that we
should have
\begin{equation}
\mu_1^2 \leq \mu_0^2 \quad\quad\quad
\mu_2^2 \leq \mu_0^2 \>.
\end{equation}
Our approach is supposed to hold up to $P_{hT} \simeq 1$ (GeV/$c$)
\cite{noi2}. Above that higher order pQCD corrections must be taken into
account, and lead to tiny variations of the values given in Eq. (\ref{par})
\cite{noi2}; however, we shall consider experiments which are expected to
produce data mainly in the low $P_{hT}$ region, and both our approach and
$\mu_{0,D}^2$values are well adequate.

We consider the $P_{hT}$ dependence of the double longitudinal spin asymmetry
\begin{equation}\label{asym}
A_{LL}(x,y,z,P_{hT}) = \frac{\int_0^{2\pi} \, d\phi_h
\, [ \, d\sigma^{\begin{array}{c}\hspace*{-0.1cm}\to\vspace*{-0.25cm}\\
\hspace*{-0.1cm}\Leftarrow\end{array}}
- d\sigma^{\begin{array}{c}\hspace*{-0.1cm}\to\vspace*{-0.25cm}\\
\hspace*{-0.1cm}\Rightarrow\end{array}} ]}
{\lambda S \int_0^{2\pi} \, d\phi_h \, [ \,
d\sigma^{\begin{array}{c}\hspace*{-0.1cm}\to\vspace*{-0.25cm}\\
\hspace*{-0.1cm}\Leftarrow\end{array}}
+ d\sigma^{\begin{array}{c}\hspace*{-0.1cm}\to\vspace*{-0.25cm}\\
\hspace*{-0.1cm}\Rightarrow\end{array}} ]} \>,
\end{equation}
and the $\cos\phi_h$ weighted asymmetry, defined as
\begin{equation}\label{wasym}
A_{LL}^{\cos\phi_h}(x,y,z,P_{hT}) = \frac{2\int_0^{2\pi} \, d\phi_h
\, [ \, d\sigma^{\begin{array}{c}\hspace*{-0.1cm}\to\vspace*{-0.25cm}\\
\hspace*{-0.1cm}\Leftarrow\end{array}}
- d\sigma^{\begin{array}{c}\hspace*{-0.1cm}\to\vspace*{-0.25cm}\\
\hspace*{-0.1cm}\Rightarrow\end{array}} ]\cos\phi_h}
{\lambda S \int_0^{2\pi} \, d\phi_h \, [ \,
d\sigma^{\begin{array}{c}\hspace*{-0.1cm}\to\vspace*{-0.25cm}\\
\hspace*{-0.1cm}\Leftarrow\end{array}}
+ d\sigma^{\begin{array}{c}\hspace*{-0.1cm}\to\vspace*{-0.25cm}\\
\hspace*{-0.1cm}\Rightarrow\end{array}} ]} \>\cdot
\end{equation}

From Eqs.~(\ref{hf1})--(\ref{hg1t}) one has
\begin{equation}
A_{LL}(x,y,z,P_{hT}) = \frac{\Delta \sigma_{LL}}{\sigma_0} ,\label{asll}
\end{equation}
with
\begin{equation}\label{all}
\Delta \sigma_{LL} = \frac{y(2-y)}{xy^2}\frac{1} {\mu_D^2 + z^2 \mu_2^2}
\exp\left(-\frac{P_{hT}^2}{\mu_D^2 + z^2 \mu_2^2}\right)
\sum_q e_q^2 \, g^q_1(x) \, D_q^h(z) \>.
\end{equation}
and
\begin{equation}\label{s0}
\sigma_0=\frac{1+(1-y)^2}{xy^2}\frac{1} {\mu_D^2 + z^2 \mu_0^2}
\exp\left(-\frac{P_{hT}^2}{\mu_D^2 + z^2 \mu_0^2}\right)
\sum_q e_q^2 \, f^q_1(x) \, D_q^h(z) \>.
\end{equation}

Analogously, Eqs.~(\ref{hf1})--(\ref{hg1t}) and (\ref{ttrpol}) give
\begin{equation}
A_{LL}^{\cos\phi_h}(x,y,z,P_{hT}) = \frac{\Delta \sigma_{LL}^{\cos\phi_h} +
\Delta \sigma_{LT}^{\cos\phi_h}}{\sigma_0} ,\label{acosw}
\end{equation}
where the contribution from the longitudinal part of the target polarization
is given by
\begin{equation}\label{sll}
\Delta \sigma_{LL}^{\cos\phi_h} = -4 \, \frac{\sqrt {1 - y}}{x y}
\frac{z \, \mu_2^2 \, P_{hT}}{{Q \, (\mu_D^2 + z^2 \mu_2^2)^2}}
\exp\left(-\frac{P_{hT}^2}{\mu_D^2 + z^2 \mu_2^2}\right)\sum_q
e_q^2 \, g^q_1(x) \, D_q^h(z),
\end{equation}
and the contribution from the transverse part of the target polarization by
\begin{equation}\label{slt}
\Delta \sigma_{LT}^{\cos\phi_h} = \frac{-2 (2-y) \sqrt{1-y}}{y}
\frac{z \, \mu_1^2 \, P_{hT}}{Q \, (\mu_D^2 + z^2 \mu_1^2)^2}
\exp\left(-\frac{P_{hT}^2}{\mu_D^2 + z^2 \mu_1^2}\right)\sum_q
e_q^2 \, g^{q}_{1T}(x) \, D_q^h(z) \>.
\end{equation}

Of course, both the numerator and denominator of Eqs. (\ref{asym}) and
({\ref{wasym}) can be integrated over some of the variables, according to the
range covered by the setups of the experiments we shall consider:
\begin{itemize}
\item {COMPASS: positive ($h^+$), all ($h$) and negative ($h^-$) hadron
production,
$Q^2 > 1.0$ (GeV/c)$^2$, $W^2 > 25$ GeV$^2$, $0.1 < x < 0.6$, $0.5 < y < 0.9$
and  $0.4 < z <0.9 $}
\item {HERMES: $\pi^+$, $\pi^0$ and $\pi^-$ production,
$Q^2 > 1.0$ (GeV/c)$^2$, $W^2 > 10$ GeV$^2$, $0.1 < x < 0.6$, $0.45 < y < 0.85$
and  $0.4 < z <0.7 $}
\item {JLab at 6 GeV: $\pi^+$, $\pi^0$ and $\pi^-$ production,
$Q^2 > 1.0$ (GeV/c)$^2$, $W^2 > 4$ GeV$^2$, $0.2 < x < 0.6$, $0.4 < y < 0.85$
and  $0.4 < z <0.7 $.}
\end{itemize}

We start by considering Eqs. (\ref{asll})--(\ref{s0}). Notice that they are
leading-twist quantities, not suppressed by any inverse power of $Q$.
Concerning the usual integrated distribution and fragmentation functions we
use the LO GRV98~\cite{grv} unpolarized and the corresponding
GRSV2000~\cite{grsv} polarized (standard scenario) DFs, and
Kretzer~\cite{kretzer} FFs. We can then compute the $P_{hT}$ dependence
of $A_{LL}$, depending on the only unknown quantity $\mu_2^2$. We plot the
results of our computations in Figs. 1 and 2, for a proton and deuteron
(+ neutron, for JLab) target, respectively.
\begin{figure}[h!]
\begin{center}
\includegraphics[width=0.6\linewidth, height=0.56\linewidth]
%{ALL_b0_pt_3x3myu_proton.eps}
{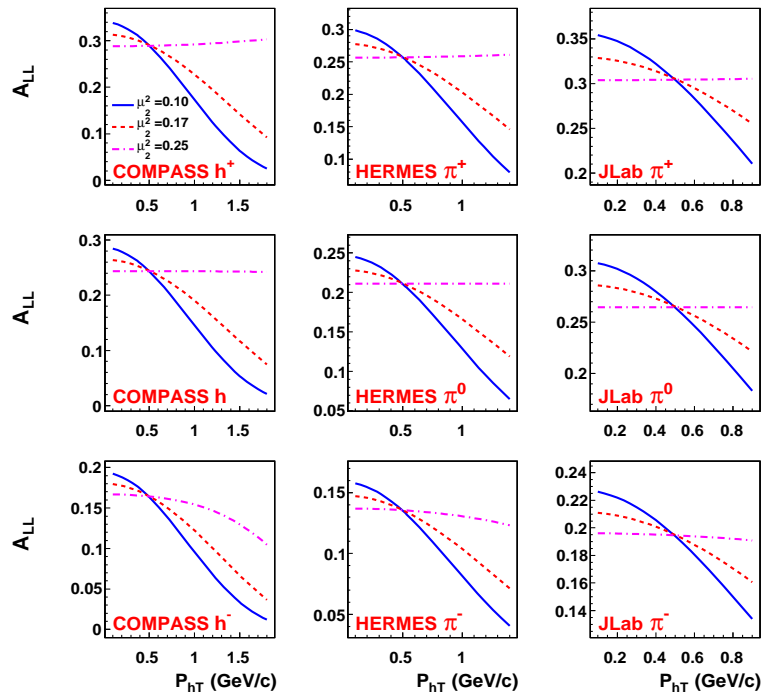} \caption{\label{fig:allptprt}{ Predicted dependence of
$A_{LL}$ on $P_{hT}$, for scattering off a proton target, with
different choices of $\mu_2^2$: 0.1 (GeV/c)$^2$ -- continuous, 0.17
(GeV/c)$^2$ -- dashed and 0.25 (GeV/c)$^2$ -- dot-dashed lines.}}
\end{center}
\end{figure}

\begin{figure}[h!]
\begin{center}
\includegraphics[width=0.6\linewidth, height=0.56\linewidth]
%{ALL_b0_pt_3x3myu_deytron.eps}
{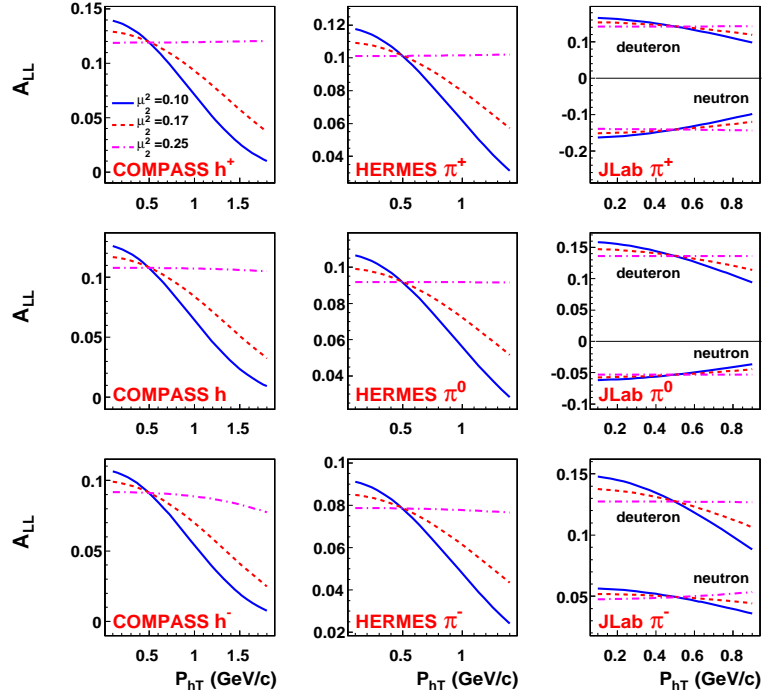} \caption{\label{fig:allptdtr}{ Predicted dependence of
$A_{LL}$ on $P_{hT}$, for scattering off a deuteron (and neutron for
JLab) target, with different choices of $\mu_2^2$: 0.1 (GeV/c)$^2$
-- continuous, 0.17 (GeV/c)$^2$ -- dashed and 0.25 (GeV/c)$^2$ --
dot-dashed lines.}}
\end{center}
\end{figure}

The results depend clearly on the relative values of $\langle
k_\perp^2 \rangle$ for the unpolarized and helicity distribution,
$\mu_0^2$ and $\mu_2^2$ respectively: $A_{LL}(P_{hT})$ is
approximately constant if $\mu_2^2 = \mu_0^2 = 0.25$ (GeV/$c)^2$,
whereas it sharply decreases with $P_{hT}$ if $\mu_2^2 < \mu_0^2$.
The trend of $A_{LL}(P_{hT})$ is thus a significant indication of
the average quark transverse motion inside unpolarized versus
longitudinally polarized nucleons. Although our numerical estimates
are based on the gaussian factorization ansatz, Eqs. (11)-(14), we
expect them to have a more general interpretation and information
content. The $P_{hT}$ dependence of $A_{LL}$ reflects, essentially,
the difference between the $k_\perp$ dependence of
$f_1^q(x,k_\perp)$ and $g_{1L}^q(x,k_\perp)$, independently of their
functional forms; the trend of
 $A_{LL}(P_{hT})$, whether constant or decreasing, reveals the behavior
 of $g_{1L}^q(x,k_\perp)/f_1^q(x,k_\perp)$ and their relative $k_\perp$
 dependence.

Similarly, we can use Eqs. (\ref{s0})--(\ref{slt}) in order to give some
estimates of $A_{LL}^{\cos\phi_h}$. Notice that
$\Delta \sigma_{LL}^{\cos\phi_h}$ and $\Delta \sigma_{LT}^{\cos\phi_h}$ are
(kinematical) higher-twist quantities, proportional to $P_{hT}/Q$; in
addition, $\Delta \sigma_{LT}^{\cos\phi_h}$ contains one unknown function,
namely $g^{q}_{1T}(x)$, related to the helicity distribution of partons inside
a transversely polarized proton. In the absence of any better guidance, we
adopt the same strategy as in Ref. \cite{kpp}. We start by noticing that,
from Eq. (\ref{dfffg3}):
\begin{equation}
g_{1T}^{q\,(1)}(x) \equiv \int d^{\,2}k_\perp \, \frac{k_\perp^2}{2M^2}
\> g^{q\perp}_{1T}(x,k_T^2) = \frac{\mu_1^2}{2M^2} \, g_{1T}^q(x)
\label{g1t1} \>.
\end{equation}
According to Refs.~\cite{tm1,mt} $g_{1T}^{q\,(1)}(x)$ is directly related
to the DF $g_2^q(x)$, which has both twist-two and higher-twist contributions,
\begin{equation}
g^q_2(x) = \frac{d}{dx}\,g^{q\,(1)}_{1T}(x) \label{gt2}.
\end{equation}

This relation, although much debated, arises from constraints imposed by
Lorentz invariance on the antiquark-target forward scattering amplitude and
the use of QCD equations of motion for quark fields~\cite{mt}. If, in
addition, one uses the Wandzura and Wilczek~\cite{ww} approximation for the
twist-two part of $g^q_2(x)$,
\begin{equation}
g_2^q(x) \simeq -g^q_1(x) + \int_x^1 dx' \,\frac{g^q_1(x')}{x'}, \label{g2ww}
\end{equation}
the following relation can be derived \cite{km},
\begin{equation}
g_{1T}^{q(1)}(x) \simeq x\int_x^1 dx'\,\frac{g^q_1(x')}{x'} \>,
\label{g11tww}
\end{equation}
which, via Eq. (\ref{g1t1}), allows to express $g_{1T}^{q}(x)$ through the
well known integrated helicity distributions.

Although such a procedure is appealing and convenient, we should stress there
are strong arguments \cite{kundu-metz}--\cite{goeke} against the validity of
the relation (\ref{gt2}). Therefore, we should consider the above expression,
Eq. (\ref{g11tww}), only as a rough model for the otherwise unknown function
$g_{1T}^{q}(x)$.

\begin{figure}[h!]
\begin{center}
\includegraphics[width=0.6\linewidth,height=0.56\linewidth]
%{ALL_pt_3x3myu_proton.eps}
{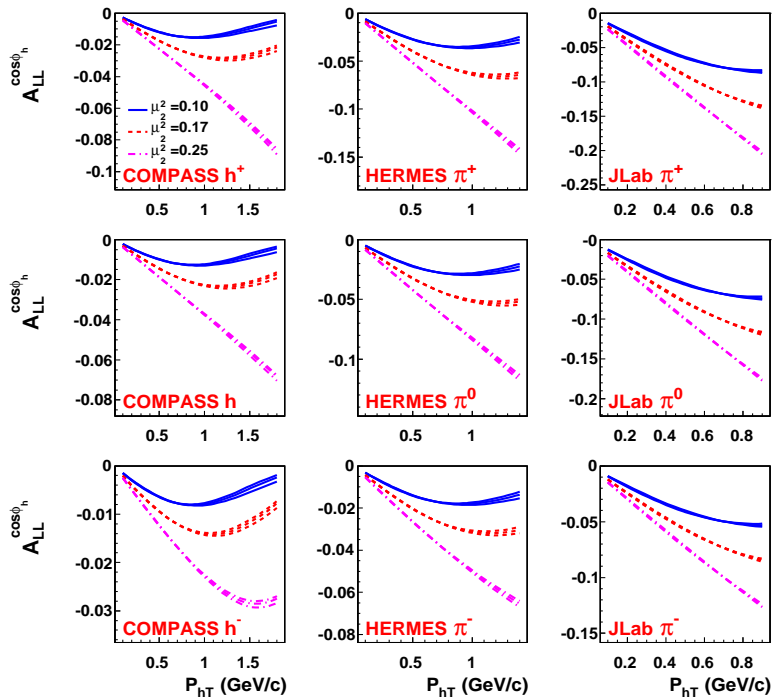} \caption{\label{fig:ptprt}{ Predicted dependence of
$A_{LL}^{\cos\phi_h}$ on $P_{hT}$ for scattering off a proton target
with different choices of $\mu_2^2$: 0.1 (GeV/c)$^2$ -- continuous,
0.17 (GeV/c)$^2$ -- dashed and 0.25 (GeV/c)$^2$ -- dot-dashed lines.
Each line splits into three almost overlapping lines corresponding,
for each value of $\mu_2^2$, to three different values of $\mu_1^2
=$ (up-down) 0.1, 0.15 and 0.2 (GeV/$c)^2$.}}
\end{center}
\end{figure}

\begin{figure}[h!]
\begin{center}
\includegraphics[width=0.6\linewidth, height=0.56\linewidth]
%{ALL_pt_3x3myu_deytron.eps}
{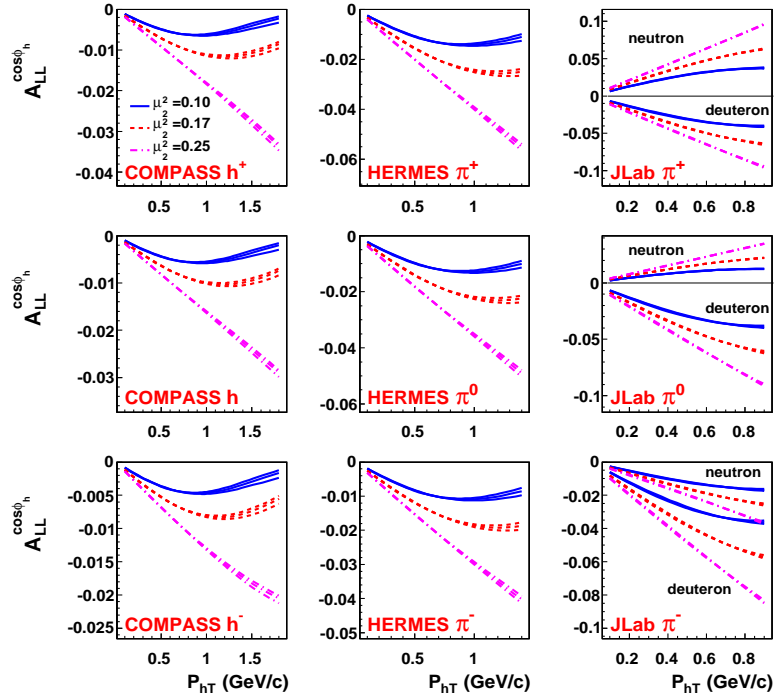} \caption{\label{fig:ptdtr}{ Predicted dependence of
$A_{LL}^{\cos\phi_h}$ on $P_{hT}$ for scattering off a deuteron (and
neutron for JLab) target with different choices of $\mu_2^2$: 0.1
(GeV/c)$^2$ -- continuous, 0.17 (GeV/c)$^2$ -- dashed and 0.25
(GeV/c)$^2$ -- dot-dashed lines. Each line splits into three almost
overlapping lines corresponding, for each value of $\mu_2^2$, to
three different values of $\mu_1^2 =$ (up-down) 0.1, 0.15 and 0.2
(GeV/$c)^2$.}}
\end{center}
\end{figure}

In Fig.~\ref{fig:ptprt} we show our predictions for
$A_{LL}^{\cos\phi_h}(P_{hT})$ as measurable by COMPASS, HERMES and JLab
collaboration experiments on a proton target. The analogous results, for
scattering off a deuteron target (and a neutron target as well, for JLab)
are shown in Fig. \ref{fig:ptdtr}. Again, we present the results for three
different choices of $\mu_2^2=$ 0.1, 0.17 and 0.25 (GeV/c)$^2$, which turn out
to be well different from each other. Instead, when varying the values of
$\mu_1^2$ our results hardly change: each line, obtained at a fixed $\mu_2^2$
value, simply splits in three almost overlapping lines (corresponding, from
up down, to  $\mu_1^2=$ 0.1, 0.15 and 0.2 (GeV/c)$^2$). This is not
surprising, as, when adopting the expression (\ref{g1t1}), there remains
little dependence on $\mu_1^2$ in Eq. (\ref{slt}). Our computations show
instead a clear strong dependence on $\mu_2^2$.

\begin{figure}[h!]
\begin{center}
\includegraphics[width=0.6\linewidth, height=0.56\linewidth]{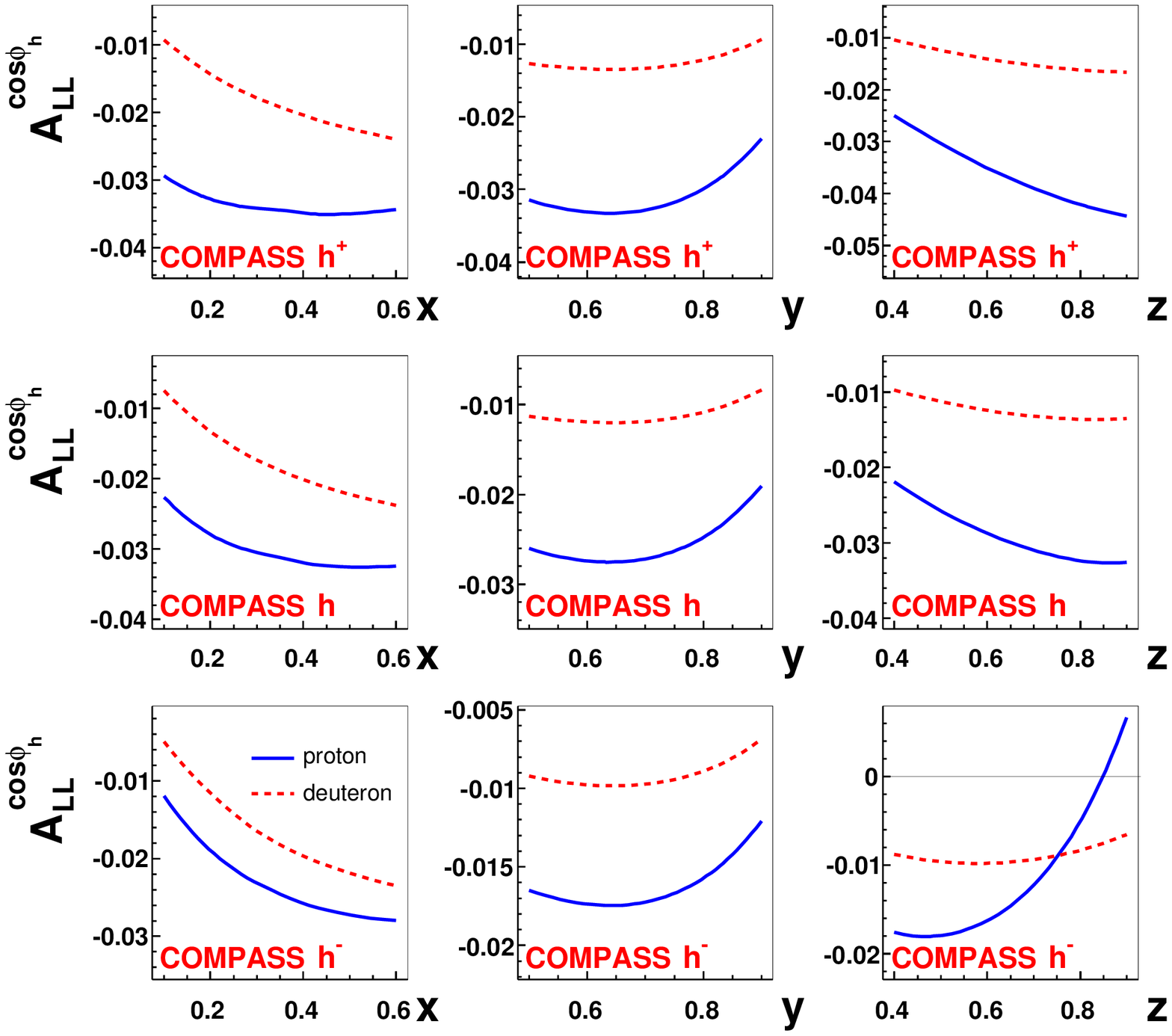}
\caption{\label{fig:cptxyz}{ Predicted dependence of $A_{LL}^{\cos\phi_h}$
on $x$, $y$ and $z$, for proton and deuteron targets, for COMPASS.}}
\end{center}
\end{figure}

\begin{figure}[h!]
\begin{center}
\includegraphics[width=0.6\linewidth, height=0.56\linewidth]{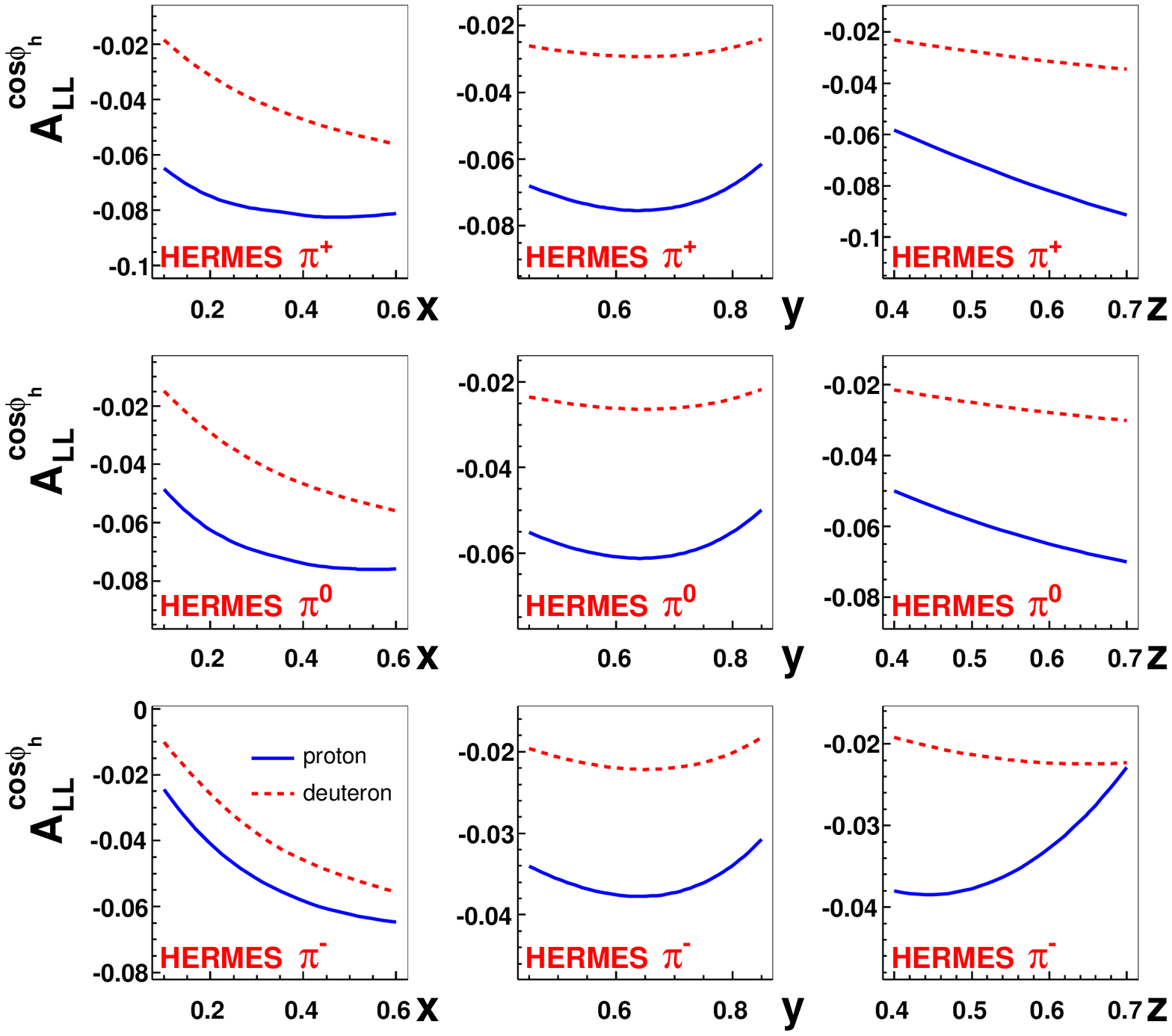}
\caption{\label{fig:hptxyz}{ Predicted dependence of $A_{LL}^{\cos\phi_h}$
on $x$, $y$ and $z$, for proton and deuteron targets, for HERMES.}}
\end{center}
\end{figure}

\begin{figure}[h!]
\begin{center}
\includegraphics[width=0.6\linewidth, height=0.56\linewidth]{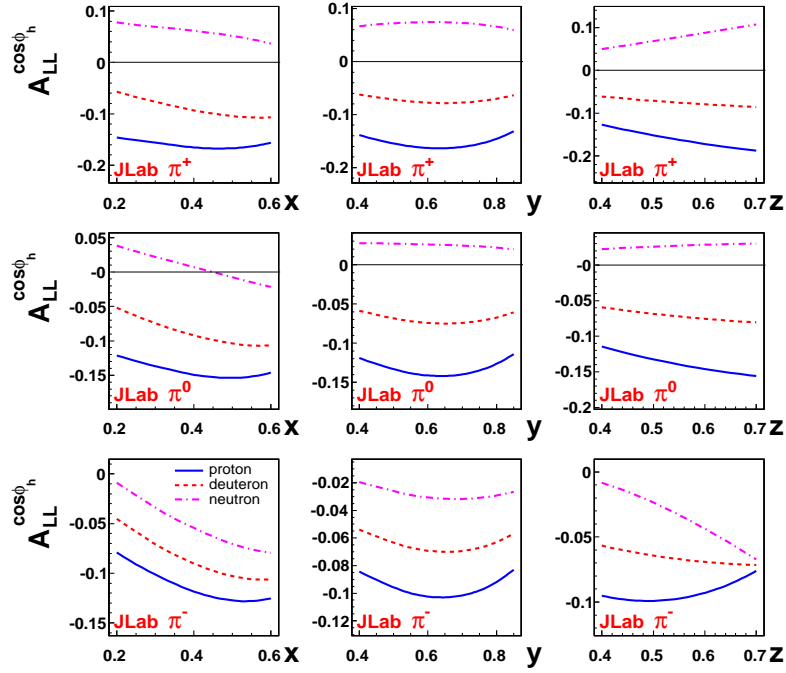}
\caption{\label{fig:jptxyz} { Predicted dependence of $A_{LL}^{\cos\phi_h}$
on $x$, $y$ and $z$, for proton, neutron and deuteron targets, for JLab.}}
\end{center}
\end{figure}

It is interesting also to compute the dependence of $A_{LL}^{\cos\phi_h}$
on each of the other single variables; for example, the $x$-dependence is
computed as
\begin{equation}\label{aptdep}
A_{LL}^{\cos\phi_h}(x) = \frac{
\, \int_{P_{hT,min}^2}^{P_{hT,max}^2}\,dP_{hT}^2\, \int dy \int dz
\, (\Delta \sigma_{LL}^{\cos\phi_h} + \Delta \sigma_{LT}^{\cos\phi_h})}
{\,\int_{P_{hT,min}^2}^{P_{hT,max}^2}\,dP_{hT}^2\, \int dy \int dz
\, \sigma_0} \>,
\end{equation}
while the $y$- and $z$-dependences are calculated in a similar way.
In Figs.~\ref{fig:cptxyz}, \ref{fig:hptxyz} and~\ref{fig:jptxyz}
we present the $x$-, $y$- and $z$-dependences of $A_{LL}^{\cos\phi_h}$
integrated over $P_{hT}$ with $P_{hT, min} = 0.5$ GeV/c and $\mu_1^2=$0.15
(GeV/c)$^2$, $\mu_2^2=$0.25 (GeV/c)$^2$ for COMPASS ($P_{hT, max} = 2$
GeV/$c$), HERMES ($P_{hT, max} = 1.5$ GeV/$c$) and JLab ($P_{hT, max} = 1$
GeV/$c$) kinematics.

\section{\label{sec:concl}Discussion and Conclusions}

We have studied the $P_{hT}$ dependence of $A_{LL}$ and $A_{LL}^{\cos\phi_h}$,
measurable in SIDIS processes by COMPASS, HERMES and JLab collaborations.
For $P_{hT}$ values up to $\sim$ 1 GeV/$c$ this dependence is entirely
generated by intrinsic motion, both of partons inside the nucleons, and of
hadrons in the parton fragmentation process \cite{anskpr,noi2}.

Within a simple factorized gaussian model for the $k_\perp$ and $p_\perp$
dependence of the distribution and fragmentation functions, it turns out
that $A_{LL}(P_{hT})$ is strongly sensitive to the relative value of
$\langle k_\perp^2 \rangle$ in unpolarized ($\mu_0^2$) and helicity
($\mu_2^2$) quark distributions: similar values, $\mu_0^2 \simeq \mu_2^2$,
would reflect into an approximately constant $A_{LL}(P_{hT})$, while
$\mu_2^2 < \mu_0^2$, would lead to a decreasing trend. Such different
behaviours are expected in general, independently of the factorized gaussian
assumption, as the shape of $A_{LL}(P_{hT})$ is essentially related to the
ratio of the $k_\perp$ dependence of $g_1^q$ and $f_1^q$. Notice, however,
that we have assumed the same constant values of $\langle k_\perp^2 \rangle$
and $\langle p_\perp^2 \rangle$ for all quark flavors; more involved choices
might lead to different behaviours. A comparison of
the quark intrinsic transverse momentum in unpolarized and longitudinally
polarized protons might give new important information concerning the spin
and orbital motion of quarks. For example, one expects that parton transverse
motion contributes to the longitudinal component of the angular momentum,
differently inside unpolarized and longitudinally polarized nucleons.

The $P_{hT}$ dependence of $A_{LL}^{\cos\phi_h}$ is not only related
to kinematical non-collinear contributions, but also to a new TMD
and spin dependent function, which gives the number density of
longitudinally polarized quarks inside a transversely polarized
nucleon. This function, $g_{1T}^{q\perp}$, induces a $\cos\phi_h$
dependence, but it is unknown; we adopted a much debated
relationship, together with the twist-two part the Wandzura-Wilczek
sum rule (and the usual Gaussian factorization), in order to link
the $x$-dependent part of $g_{1T}^{q\perp}$ to the integrated
helicity distributions. Within such an approach, it turns out that
also $A_{LL}^{\cos\phi_h}(P_{hT})$ has a strong dependence on
$\mu_2^2$ alone, thus giving further information on the average
transverse motion of quarks inside a longitudinally polarized
proton.

We conclude by noticing, as it was done in Ref. \cite{kpp}, that the exact
$k_\perp$ dependence of the distribution functions $f_1^q(x,k_\perp)$,
$g_{1L}^q(x,k_\perp)$ and $g_{1T}^{q\perp}(x, k_\perp)$ is crucial when
considering the general positivity bounds of Ref. \cite{bacchetta}, which
involve in one inequality the three previous functions and the Sivers
function. The $k_\perp$ dependence might play an essential role in fulfilling
the inequality, and a check of its validity is a fundamental test for the
self consistency of the LO QCD description of SIDIS processes.

\section*{Acknowledgements}
This research is part of the EU Integrated Infrastructure Initiative
HadronPhysics project, under contract number RII3-CT-2004-506078.
A.E is also supported by Grants RFBR 06-02-16215 and RF MSE
RNP.2.2.2.2.6546. M.A. acknowledges partial support by MIUR under
contract 2004021808\_009.


\begin{thebibliography}{99}

\bibitem{herm}
  A.~Airapetian {\it et al.}  [HERMES Collaboration],
  Phys.\ Rev.\ Lett.\  {\bf 94}, 012002 (2005)

\bibitem{comp}
  V.~Y.~Alexakhin {\it et al.}  [COMPASS Collaboration],
  Phys.\ Rev.\ Lett.\  {\bf 94}, 202002 (2005).

\bibitem{jlab}
  H.~Avakian {\it et al.}  [CLAS Collaboration],
  Phys.\ Rev.\ D {\bf 69}, 112004 (2004).

  H.~Avakian, P.~Bosted, V.~Burkert and L.~Elouadrhiri  [CLAS Collaboration],
  AIP Conf.\ Proc.\  {\bf 792}, 945 (2005),e-Print Archive: nucl-ex/0510032.

\bibitem{cahn}
  R.~N.~Cahn,
  Phys.\ Lett.\ B {\bf 78}, 269 (1978);
  Phys.\ Rev.\ D {\bf 40}, 3107 (1989).

\bibitem{anskpr}
  M.~Anselmino, M.~Boglione, U.~D'Alesio, A.~Kotzinian, F.~Murgia and A.~Prokudin,
  Phys.\ Rev.\ D {\bf 71}, 074006 (2005);
  Phys.\ Rev.\ D {\bf 72}, 094007 (2005)
  [Erratum-ibid.\ D {\bf 72}, 099903 (2005)].

\bibitem{noi2}
  M.~Anselmino, M.~Boglione, A. Prokudin and C. T\"urk,
  e-Print Archive: hep-ph/0606286.

\bibitem{ko}
  A.~Kotzinian, Nucl. Phys. {\bf B441}, 234 (1995).

\bibitem{mt}
  P.J.~Mulders and R.D.~Tangerman, Nucl. Phys. {\bf B461}, 197
  (1996), Nucl. Phys. {\bf B484}, 538 (1997), Erratum.

\bibitem{kpp}
  A.~Kotzinian, B.~Parsamyan and A.~Prokudin,
  %``Predictions for double spin asymmetry A(LT) in semi inclusive DIS,''
  Phys.\ Rev.\ D {\bf 73}, 114017 (2006)
  [arXiv:hep-ph/0603194].
  %%CITATION = HEP-PH 0603194;%%

\bibitem{jmy}
  X.~Ji, J.P.~Ma and F. Yuan, Phys.\ Rev.\ {\bf D71}, 034005 (2005).

\bibitem{noi}
  M.~Anselmino, M.~Boglione, U.~D'Alesio, E.~Leader, S.~Melis and F.~Murgia,
  Phys.\ Rev.\ D {\bf 73} (2006) 014020.

\bibitem{grv}
  M.~Gluck, E.~Reya and A.~Vogt,
  Eur.\ Phys.\ J.\ C {\bf 5}, 461 (1998).

\bibitem{grsv}
  M.~Gluck, E.~Reya, M.~Stratmann and W.~Vogelsang,
  Phys.\ Rev.\ D {\bf 63}, 094005 (2001).

\bibitem{kretzer}
  S.~Kretzer,
  Phys.\ Rev.\ D {\bf 62}, 054001 (2000).

\bibitem{tm1}
  R.D.~Tangerman and P.J.~Mulders, e-Print Archive: hep-ph/9408305.

\bibitem{ww}
  S.~Wandzura and F.~Wilczek, Phys. Lett. {\bf B72}, 195
  (1977).

\bibitem{km} A.~Kotzinian and P.J.~Mulders,
  Phys. Rev. {\bf D54}, 1229 (1996).

\bibitem{kundu-metz}
  R.~Kundu and A.~Metz,
  Phys.\ Rev.\ D {\bf 65} (2001) 014009.

\bibitem{schlegel-metz}
  M.~Schlegel and A.~Metz, e-Print Archive: hep-ph/0406289.

\bibitem{goeke}
  K.~Goeke, A.~Metz, P.~V.~Pobylitsa and M.~V.~Polyakov,
  Phys.\ Lett.\ B {\bf 567} (2003) 27.

\bibitem{bacchetta}
  A.~Bacchetta, M.~Boglione, A.~Henneman and P.~J.~Mulders,
  Phys.\ Rev.\ Lett.\  {\bf 85}, 712 (2000).

\end{thebibliography}
\end{document}